\documentstyle[preprint,aps,psfig]{revtex}
\begin{document}
\draft
\preprint{}
\title{Vacuum polarization effects in hyperon rich dense matter --\\
a nonperturbative treatment}
\author{Amruta Mishra$^{\dagger}$, P.K. Panda\footnote[1]{
email: panda@ift.unesp.br}$^\ddagger$, 
W. Greiner$^\S$}
\address{$^\dagger$ Institute for Plasma Research,
Gandhinagar - 382428, India} 
\address{$^\ddagger$ Instituto de F\'{\i}sica Te\'orica,
Universidade Estadual Paulista,\\ Rua Pamplona 145, 01405-900 S\~ao Paulo -
SP, Brazil}
\address {$^\S$Institut f\"ur Theoretische Physik, 
J.W. Goethe Universit\"at, Robert Mayer-Stra{\ss}e 10,\\
Postfach 11 19 32, D-60054 Frankfurt/Main, Germany}
\maketitle
\begin{abstract}
We derive the equation of state (EOS) for electrically charge neutral
dense matter using Quantum Hadrodynamics (QHD) model.
This is carried out in a nonperturbative manner
including quantum corrections for baryons 
through a realignment of vacuum with baryon-antibaryon
condensates. This yields the results of relativistic
Hartree approximation of summing over baryonic tadpole diagrams.
The quantum corrections from the scalar meson 
is also taken into account in a similar way. This leads to a 
softening of the equation of state for the hyperonic matter. 
The formalism also allows us to do a self consistent calculation of
the in-medium sigma meson mass. The effects of such quantum corrections
on the composition of charge neutral dense matter is
considered. The effect of the resulting EOS
on the structure of the neutron stars is also studied.
\end{abstract}
\pacs{PACS number: 21.65.+f,14.20.J,13.75.E}
\def\bfm#1{\mbox{\boldmath $#1$}}
\section {Introduction}

The study  of hot and dense matter is an interesting and important
problem in the strong interaction physics, which has relevance
in the context of relativistic heavy ion collision experiments,
as well as to study compact stellar objects such as neutron stars.
It is thus important to understand the strongly interacting many body
systems.

Quantum Hadrodynamics (QHD) is a general formalism for studying
nuclear many body problem \cite {walecka}. This has extensively been 
used to study nuclear matter at zero and finite temperatures, as well as
to describe the properties of finite nuclei. In the Walecka model
(QHD-I), the nucleons interact through the scalar ($\sigma$)
and vector ($\omega$) mesons. This was first studied neglecting
the Dirac sea, i.e., using a mean field approach, and is known as the
`no-sea' approximation.
To include the sea effects, one does a self consistent sum of tadpole 
diagrams for the baryon propagator \cite{chin}. There have also been 
attempts to include corrections up to two-loops \cite{twoloop}. However, 
the couplings involved here are too large (of the order of 10), 
thus making the theory
unstable against a perturbative loop expansion. This necessitates
non-perturbative techniques to be developed to consider nuclear
many-body problems.

We had earlier considered \cite{mishra} a nonperturbative treatment
for studying nuclear matter including vacuum polarization effects.
The approximation scheme uses a squeezed coherent type of construction
for the ground state, which amounts to an explicit vacuum realignment
with baryon-antibaryon condensates. The properties of nuclear matter
as arising from such a ground state were examined. Such a vacuum
was then generalized to include the quantum effects from meson fields
through scalar meson condensates. This gave rise to a softer equation
of state with a lower value for the compressibility for nuclear matter.
The density dependent sigma meson mass was also calculated including
the quantum effects.

It is believed that the formation of hyperons is a dominant mechanism
at high densities and temperatures, which  contributes to the
neutron star matter and influences the equation of state (EOS)
\cite{gled,prakash}. 
These particles ($\Lambda$, $\Sigma^{\pm,0}$,
$\Xi^{-,0}$) may be excited through strangeness violating
weak processes. 
There have also been attempts to include the
contribution of the filled negative sea of hyperons to the energy 
of nuclear matter at one-loop level through summing over the
baryonic tadpole diagrams. This is the relativistic Hartree
approximation (RHA). 
The contribution arising from two-loop diagrams has also
been considered \cite{ellis}, 
which showed a strong degree of cancellation between one- and
two-loop contributions giving a negligible effect. 
However, this conclusion was strongly dependent on the
rather uncertain values of the hyperon coupling constants.

In the present work, the nonperturbative formalism described 
in ref. \cite{mishra}, is generalized to include the strange baryons and EOS
for dense matter is derived taking into account the vaccum
fluctuation effects from the baryon and scalar meson sectors
within the QHD model. The effect of such quantum corrections to the EOS
on the structure of the neutron stars is studied. We might note here that
hyperon stars have been treated recently within the Brueckner-Bethe-Goldstone
theory, including three body nucleon interactions \cite{baldo} which gives
rise to a softer EOS.

We organize the paper as follows. In section II, we derive the
EOS for dense matter including the quantum
effects arising from the baryons and $\sigma$ mesons in a nonperturbative
manner with an explicit construction for the ground state within 
the QHD model. We also consider the effect of 
such EOS on neutron star strucure.  In section III, we discuss in detail
the numerical results obtained in the present work. 
Finally, in section IV we summarise the
results and discuss possible outlook.

\section {Quantum vacuum in hyperon matter}

We start with the Lagrangian density for QHD model of baryons
interacting via exchange of scalar $\sigma$, vector $\omega$
and isovector vector $\rho$ mesons given as \cite{walecka}

\begin{eqnarray}
{\cal L}&=&\sum_B\bar \psi_B (i\gamma^\mu \partial_\mu
-M_B-g_{\sigma B}\sigma-g_{\omega B}\gamma^\mu \omega_\mu
-\frac {1}{2}g_{\rho B}\gamma_\mu {\bfm \tau}_B \cdot 
{\bfm \rho}^\mu)\psi_B
+\frac{1}{2}\partial^\mu\sigma
\partial_\mu\sigma-\frac{1}{2} m_\sigma ^2 \sigma^2
\nonumber \\ 
& + & \frac {1}{2} m_ \omega^2 \omega^\mu \omega_\mu
-\frac {1}{4}\omega^{\mu \nu}\omega_{\mu \nu}
+\frac {1}{2} m_\rho^2 {\bfm \rho}^\mu \cdot {\bfm \rho}_\mu
-\frac {1}{4} {\bfm \rho}^{\mu \nu}{\bfm \rho}_{\mu \nu}.
\end{eqnarray}
In the above, $\psi_B$, $\sigma$, $\omega_\mu$ and ${\bfm \rho}^\mu$
are the fields for the baryons including hyperons
(B=p, n, $\Lambda$, $\Sigma^{\pm,0}$, $\Xi^{-,0}$), $\sigma$, $\omega$ 
and $\rho$ -mesons
with masses $M_B$, $m_\sigma$,$m_\omega$ and $m_\rho$, respectively. 

We shall be using the mean field approximation for the meson fields. This
amounts to taking meson fields as classical fields.
For an uniform and static matter
within this approximation, only the time-like components of the vector fields
and the isospin-3 component of the $\rho$ meson field have nonvanishing
values. The mean meson fields are denoted by $\sigma_0$, $\omega_0$ and
$\rho_{03}$. However, we shall treat the fermion fields as quantum fields.
The Hamiltonian density can then be written as
\begin{equation}
{\cal H}={\cal H}_B+{\cal H}_\sigma+{\cal H}_\omega+{\cal H}_\rho
\end{equation}
with
\begin{mathletters}
\begin{equation}
{\cal H}_B=\sum_B\psi_B ^\dagger(-i {\bbox \alpha} \cdot {\bbox \nabla}
+\beta M_B)\psi_B + \sum_B g_{\sigma B}~\sigma_0\bar \psi_B\psi_B
\end{equation}
\begin{equation}
{\cal H}_\sigma= \frac{1}{2} m_\sigma ^2 \sigma_0^2
\label{lwsg}
\end{equation}
\begin{equation}
{\cal H}_\omega= \sum_B g_{\omega B}~\omega_0 \psi_B ^\dagger \psi_B
-\frac{1}{2} m_\omega ^2 \omega_0^2
\end{equation}
\begin{equation}
{\cal H}_\rho= \sum_B g_{\rho B}~I_{3B} \rho_{03} \psi_B ^\dagger \psi_B
-\frac{1}{2} m_\rho ^2 \rho_{03}^2.
\end{equation}
\end{mathletters}
The equal-time quantization condition for the baryons is given as
\begin{equation}
[\psi _{B\alpha}({\bf x},t),\psi^\dagger _{B'\beta} ({\bf y},t)]_{+}
=\delta _{BB'} \delta _{\alpha \beta}\delta({\bf x} -{\bf y}),
\end{equation}
where $\alpha$ and $\beta$ refer to the spin indices.
We may now write down the field expansion for the baryon field
of type $B$, $\psi_B$ at time $t=0$ as given by \cite{prd}
\begin{equation}
\psi_B({\bf x})=\frac {1}{(2\pi)^{3/2}}\int \left[U_{Br}({\bf k})c_{BIr}
({\bf k})+V_{Bs}(-{\bf k})\tilde c_{BIs}(-{\bf k})\right] e^{i{\bf k}
\cdot {\bf x}} d{\bf k},
\end{equation}
with $c_{BIr}$ and $\tilde c_{BIr}$ as the particle  annihilation 
and antiparticle creation operators of baryon of type $B$,
with spin $r$. 

In the above, $U_{Br}$ and $V_{Bs}$ are given by
\begin{equation}
U_{Br}({\bf k})=\left( \begin{array}{c}\cos\frac{\chi_B({\bf k})}{2}
\\ {\bbox \sigma} \cdot\hat k\sin\frac{\chi_B({\bf k})}{2}
\end{array}\right)u_{Ir} ;\quad V_{Bs}(-{\bf k})=\left(
\begin{array}{c}-{\bbox \sigma} \cdot\hat k\sin\frac{\chi_B({\bf k})}{2}
\\ \cos\frac{\chi_B({\bf k})}{2}\end{array}\right)v_{Is}~.
\end{equation}
For free massive fields $\cos\chi_B({\bf k})=M_B/\epsilon_B({\bf k})$ and
$\sin\chi_B({\bf k})=|{\bf k}|/\epsilon_B({\bf k}),$ with 
$\epsilon_B({\bf k})=\sqrt{{\bf k}^2 + M_B^2}$.
The perturbative vacuum, say $\mid vac\rangle$, is defined through
$c_{BIr}({\bf k})\mid vac\rangle=0$ and $\tilde c_{BIr}^\dagger({\bf k})\mid 
vac \rangle=0$.

\subsection {Quantum corrections from baryon sector}

We first use the mean-field approximation for the meson  fields,
and retain the quantum nature of the fermion fields \cite{gn}.
This amounts to taking meson fields as constant classical
fields with translational invariance for nuclear matter.
To include the vacuum polarization effects from the baryon sector, 
we shall now consider a trial state with baryon-antibaryon condensates 
\cite {mishra}, now generalised to include the hyperons. 
We thus explicitly take the ansatz for the above state as 
\begin{equation}
|vac'\rangle = \exp \Big[
\int d{\bf k} ~ \sum_B f_B({\bf k})~{c_{BIr}^\dagger ({\bf k})}~a_{rs}
\tilde c _{BIs} (-{\bf k})-h.c. \Big]|vac\rangle 
\equiv  U_F|vac\rangle~.
\label{barcon}
\end{equation}
Here $a_{rs}=u_{Ir}^\dagger({\bbox \sigma} \cdot \hat k)v_{Is}$ 
and $f_B({\bf k})$ is a trial function associated with baryon-antibaryon 
condensates of type $B$. We note that with the above transformation
the operators corresponding to $\mid vac'\rangle$ are related to the
operators corresponding to $|vac\rangle$ through a Bogoliubov transformation
\begin{equation}
\left(
\begin{array}{c} d_{I}({\bf k})\\{\tilde d}_{I}(-{\bf k})
\end{array}
\right)
=\left(
\begin{array}{cc}
\cos f_B({\bf k}) & -{\bbox\sigma} \cdot \hat k\sin f_B({\bf k})
\\ {\bbox\sigma} \cdot \hat k \sin f_B({\bf k}) & \cos f_B({\bf k})
\end{array}
\right)
\left(
\begin{array}{c}
c_I({\bf k})\\ {\tilde c}_I(-{\bf k})
\end{array}
\right ).
\label{bogf}
\end{equation}

We then use the method of thermofield dynamics \cite{tfd}
developed by Umezawa to construct the ground state for nuclear matter. 
We generalise the state with baryon-antibaryon condensates
as given by (\ref{barcon}) to finite temperature and density as
\cite{mishra}
\begin{equation}
|F(\beta)\rangle=U(\beta)|vac'\rangle\equiv U(\beta)U_F|vac\rangle.
\end{equation}
The temperature-dependent unitary operator $U(\beta)$ is given as
\cite{tfd}
\begin{equation}
U(\beta) =\exp{(A^\dagger (\beta)-A(\beta))},
\end{equation}
with 
\begin{equation}
A^\dagger(\beta)= \int d{\bf k} ~ \sum_B \bigg[\theta_{B-}({\bf k},\beta)~
d_{BIr}^\dagger ({\bf k})~{\underline d}_{BIr}^\dagger(-{\bf k})
+\theta_{B+}({\bf k},\beta)~
\tilde d_{BIr}({\bf k})~\tilde {\underline d}_{BIr}(-{\bf k})\bigg].
\end{equation}
The underlined operators correspond to the
doubling of the Hilbert space that arises in thermofield dynamics method
\cite {tfd}.
We shall determine the condensate function $f_B({\bf k})$,
and the functions $\theta_{B-}({\bf k},\beta)$ and $\theta_{B+}({\bf k},\beta)$ 
of the thermal vacuum through minimisation of the thermodynamic potential.
The thermodynamic potential is then given as
\begin{equation}
\Omega=\epsilon-\frac{1}{\beta}{\cal S}-\sum_B\mu_B \rho_B,
\label{thermpot}
\end{equation}
with the  energy density as
$\epsilon\equiv\langle {\cal H}\rangle_\beta$, and the 
the entropy density as
\begin{eqnarray}
{\cal S}&=&-\sum_B\gamma_B (2\pi)^{-3} \int d {\bf k} \Bigg[
\sin^2 \! \theta_{B-} \ln(\sin^2\!\theta_{B-})+
\cos^2 \! \theta_{B-} \ln(\cos^2\!\theta_{B-})\nonumber\\
&+&
\sin^2 \! \theta_{B+} \ln(\sin^2\!\theta_{B+})+
\cos^2 \! \theta_{B+} \ln(\cos^2\!\theta_{B+})
\Bigg]+{\cal S}_\sigma +{\cal S}_\omega +{\cal S}_\rho.
\label{entrf}
\end{eqnarray}
The baryon density corresponding to baryon $B$ is
\begin{equation}
\rho_B=\gamma_B (2\pi)^{- 3}\int d {\bf k} 
(\cos^2 \theta_{B+} +\sin^2 \theta _{B-})
\end{equation}
and $\mu_B$ is the chemical potential of the corresponding baryon.
In the above, $\gamma_B$  is the spin degeneracy factor. 
Further, ${\cal S}_\sigma$, ${\cal S}_\omega$  and ${\cal S}_\rho$ 
are the contributions to the entropy density from
$\sigma$, $\omega$, and $\rho$ mesons, respectively.

Extremising the thermodynamic potential $\Omega$ with respect
to the condensate function $f_B({\bf k})$ and the functions 
$\theta_{B\mp}$ yields
\begin{equation}
\tan 2 f_B({\bf k})=\frac{g_{\sigma B}~\sigma_0 |{\bf k}|}{\epsilon_B
({\bf k})^2+ M_B g_{\sigma B}~\sigma_0}~,
\end{equation}
and
\begin{equation}
\sin^2 \theta _{B\mp}=\frac{1}{\exp(\beta(\epsilon_B^*(k)\mp \mu_B^{*})) +1}~,
\end{equation}
with $\epsilon_B^*(k)=(k^2+{M_B^*}^2)^{1/2}$ 
as the effective energy density, defined through an effective
baryon mass, $M_B^{*}=M_B+g_{\sigma B}~\sigma_0$. Further,
the effective chemical potential, $\mu_B^*$ is given as
\begin{equation}
\mu_B^{*}
=\mu_B -g_{\omega B}~ \omega_0-g_{\rho B} I_{3B} \rho_{03}.
\label {mustr}
\end{equation}

To obtain the equation of state (EOS) for the nuclear matter at zero
temperature, we subtract out the pure vacuum contribution
($f_B=0$,$\;\;$$\theta_{B-}=0$ part) and renormalise the energy density
by adding the counter terms \cite{chin}. This yields the expressions
for the energy density and pressure, p (=$-\Omega$) as

\begin{equation}
\epsilon=\sum_B \frac {\gamma_B}{2\pi^2} \int _0 ^{k_{FB}} 
(k^2+{M_B^*}^2)^{1/2}
k^2 dk +\Delta \epsilon_{ren} +\frac {1}{2} {m_\sigma}^2\sigma_0^2
+\frac {1}{2} {m_\omega}^2 \omega_0^2
+\frac {1}{2} {m_\rho}^2 \rho_{03}^2,
\end{equation}
and
\begin{equation}
p= \frac {1}{3}\sum_B \frac {\gamma_B}{2\pi^2} \int _0 ^{k_{FB}} 
\frac {k^4}{(k^2+{M_B^*}^2)^{1/2}}
dk -\Delta \epsilon_{ren} -\frac {1}{2} {m_\sigma}^2\sigma_0^2
+\frac {1}{2} {m_\omega}^2 \omega_0^2
+\frac {1}{2} {m_\rho}^2 \rho_{03}^2~.
\end{equation}

In the above, $\Delta\epsilon_{ren}$ arises from the filled Dirac
sea of baryons and is given as

\begin{eqnarray}
\Delta \epsilon_{ren}&=& -\sum_B\frac{\gamma_B}{16 \pi^2}\bigg[
{M_B^*}^4 \ln \bigl(\frac{M_B^{*}}{M_B}\bigr)+M_B^3 (M_B-{M_B^*})
-\frac{7}{2}M_B^2  (M_B-{M_B^*})^{2}\nonumber\\
 &+& \frac{13}{3} M_B  (M_B-{M_B^*})^3 
-\frac{25}{12} (M_B-{M_B^*})^4\bigg].
\label{rhf}
\end{eqnarray}

The energy density and pressure are now finite and given  in terms
of the meson fields, $\sigma_0$, $\omega_0$ and $\rho_{03}$. These
mean fields are obtained through the extremisation of the thermodynamic 
potential, $\Omega$, and satisfy the self consistency conditions

\begin{mathletters}
\begin{equation}
m_\sigma^2 \sigma_0=-\sum_B g_{\sigma B} \frac {\gamma_B}{2\pi^2}
\int_0^{k_{FB}} \frac {{M_B^*}k^2 dk}{(k^2+{M_B^*}^2)^{1/2}}-
\frac {d \Delta\epsilon_{ren}}{d\sigma_0}~,
\label {sg0}
\end{equation}
\begin{equation}
m_\omega ^2 \omega_0=\sum_B g_{\omega B} \rho_B~,
\end{equation}
and,
\begin{equation}
m_\rho ^2 \rho_{03}=\sum_B g_{\omega B} I_{3B} \rho_B~.
\end{equation}
\label {selfcons}
\end{mathletters}

The expressions for the energy density and pressure are given
in terms of the fermi momenta, $k_{FB}$, which are as yet
unknown. They are given in terms of the chemical potentials,
$\mu_B$ as $k_{FB}=({\mu_B^*}^2-{M_B^*}^2)^{1/2}$, for $\mu_B^*>M_B^*$
and zero otherwise. 
Hence using (\ref{mustr}), the threshold condition for appearance
of the baryon of type $B$ is given as
\begin{equation}
\mu_B^* \equiv \mu_B -g_{\omega B}~ \omega_0-g_{\rho B} I_{3B} \rho_{03} 
\ge M_B+g_{\sigma B} \sigma_0.
\label {thresh}
\end{equation}
These, in turn, can be written in terms of the two independent 
chemical potentials,
$\mu$ and $\mu_E$ (corresponding to the baryon and electric charge
conservation) as
\begin{equation}
\mu_B=\mu+q_B\mu_E,
\end{equation}
where $q_B$ is the electric charge of baryon B.

The quantities $\mu$ and $\mu_E$ are determined from the
baryon charge conservation

\begin {equation}
\rho=\sum_B \rho_B
\label{chrgb}
\end{equation}
and the electric charge neutrality condition (since the
matter inside the neutron star is electrically charge neutral) 
\begin{equation}
\sum_{i=B,e,\mu} q_i \rho_i=0
\label{chrg0}
\end{equation}
where $\rho_i=\gamma_i k_{Fi}^3/6\pi^2$, with $\gamma_i$ as the
spin degeneracy factor and $k_{Fi}$ as the fermi momentum for
particle of species $i$.

The equations (\ref{selfcons}) to eqn (\ref{chrg0}) comprise a coupled
set of equations to determine the quantities $\sigma_0$, $\omega_0$,
$\rho_{03}$, $\mu_E$, $\mu$, and, $k_{Fi}$'s ($i$=e, $\mu$, $B$).

We note that the self consistency condition for the $\sigma$
field as well as the energy density as obtained through an
explicit construct of a state with baryon-antibaryon
condensates, including the hyperons, are identical to those 
obtained through summing the tadpole diagrams for the
baryon propagator in the relativistic Hartree approximation
\cite {chin,prakash}.

\subsection {Quantum corrections from scalar mesons}

We next consider the quantum corrections due to the
scalar mesons. It arises from a vacuum realignment
with sigma condensates \cite{mishra,ph4} meaning thereby $\sigma$-field 
is not a classical one, but a quantum field.
As will be seen later, a quartic term in the sigma field
would favour such condensates. Self-interactions
of scalar fields with cubic and quartic terms have been considered 
earlier \cite{boguta}
in the no-sea approximation \cite{finnl} as well as including the
quantum  corrections arising from the sigma fields \cite{walecka,glen,fox}.
They may be regarded as mediating three- and four-body 
interactions between the baryons.
The best fits to incompressibility in nuclear matter,
single-particle spectra and properties of deformed nuclei are
achieved with a negative value for the quartic coupling in the
sigma field. However, with such a negative coupling the energy spectrum
of the theory becomes unbounded from below \cite{furnstahl}
for large $\sigma$ and hence it is impossible to study
excited spectra or to include vacuum polarization effects
\cite {mishra}.

Including a quartic scalar self-interaction, eq. (\ref{lwsg}) is modified to 
\begin{equation}
{\cal H}_\sigma= \frac{1}{2}{\dot \sigma} ^2
+\frac {1}{2} ({\bbox \nabla} \sigma)^2
+\frac{1}{2} m_\sigma ^2 \sigma^2+\lambda \sigma^4,
\label{lwsgm}
\end{equation}
with $m_\sigma$ and $\lambda$ being the bare mass and coupling 
constant respectively. 
We may expand the $\sigma$ field in terms of creation and 
annihilation operators at time $t=0$ as
\begin{equation}
\sigma ({\bf x},0)={1\over{(2 \pi)^{3/2}} }\int{{d{\bf k}\over{\sqrt
{2 \omega ({\bf k})}}}\left(a({\bf k})+
a^\dagger(-{\bf k})\right)e^{i{\bf k}\cdot{\bf x}}}.
\label{expan}
\end{equation}
In the above, $\omega ({\bf k})$ is an arbitrary function which 
for free fields is given by $\omega ({\bf k})=\sqrt{{\bf k}^2+m_\sigma^2}$ 
and the perturbative vacuum is defined corresponding to this basis through 
$a\mid vac\rangle=0$. 

As seen earlier a realignment of the
ground state from $\mid vac\rangle$ to $\mid vac'\rangle$
with baryon condensates amounts to including quantum effects arising
from the baryons. We shall adopt a similar procedure now to calculate
the quantum corrections arising from the $\sigma$-field
\cite {mishra}. We thus modify
the ansatz for the trial ground state as given by (\ref{barcon}) to 
include $\sigma$ condensates as \cite{mishra,ph4}
\begin{equation}
|\Omega\rangle=U_\sigma U_F|vac\rangle,
\label{barsgcon}
\end{equation}
with
$U_\sigma=U_{II}U_{I}$,
where $U_{i}=\exp(B_i^\dagger~-~B_i),\,(i=I,II)$. Explicitly the
$B_{i}$ are given as
\begin{equation}
B_I^\dagger=\int {d{\bf k} \sqrt{\omega ({\bf k})\over 2}
f_\sigma({\bf k}) a^\dagger({\bf k})},\;\;\;\;
{B_{II}}^\dagger={1\over 2}\int d{\bf k} g({\bf k}){a'}^\dagger({\bf k})
{a'}^\dagger(-{\bf k}). 
\end{equation}
In the above, $a'({\bf k})=U_I a({\bf k}) U_I^{-1}=a({\bf k})-
\sqrt{\frac{\omega ({\bf k})}{2}}f_\sigma({\bf k})$ corresponds to a
shifted field operator associated with the coherent state \cite{mishra,ph4}.
Thus in this construct for the ground state we have two
functions $f_\sigma({\bf k})$ and $g({\bf k})$ which will be determined through
minimisation of energy density. Further, since $\mid \Omega\rangle$ contains 
an arbitrary number of $a'^{\dagger}$ quanta, $a'\mid \Omega\rangle\,\not=\,0$. 
However, we can define the basis $b({\bf k})$, $b^\dagger({\bf k})$ 
corresponding to $\mid \Omega\rangle$ through the Bogoliubov transformation as
\begin{equation}
b({\bf k}) =  U _{II} a'({\bf k}) U_{II}^{-1}=
(\cosh\! g )a'({\bf k}) -(\sinh\! g) a'^\dagger(-{\bf k})~.
\label{gep7}
\end{equation}
It is easy to check that $b({\bf k})\mid \Omega\rangle=0 $. Further, to preserve 
translational invariance $f_\sigma({\bf k})$ has to be  proportional to $\delta
({\bf k})$ and  we take $f_\sigma({\bf k})=\sigma _{0} (2\pi)^{3/2}
\delta ({\bf k})$. $\sigma_0$ will correspond to a
classical field of the conventional approach \cite{ph4}.
We next calculate the expectation value of the Hamiltonian density.

Using the transformations (\ref{gep7}) it is easy to evaluate that
\begin{equation}
\langle \Omega\mid \sigma \mid \Omega\rangle=\sigma _{0},\;\;\;
\langle \Omega\mid \sigma^2 \mid \Omega\rangle={\sigma _{0}}^{2}+I,
\label{gep9}
\end{equation}
where
\begin{equation}
I={1 \over (2 \pi)^3}\int{{d{\bf k} \over {2\;\omega (k)}}
(\cosh \!2g +\sinh\!2g)}.
\label{gep9c}
\end{equation}
Using equations (\ref{lwsgm}), (\ref{gep9}), the energy density of 
${\cal H}_\sigma$ with respect to the trial state becomes \cite{ph4}
\begin{eqnarray}
\epsilon_\sigma\equiv\langle \Omega\mid{\cal H}_\sigma\mid \Omega\rangle &=&
\frac{1}{2}{1 \over {(2 \pi)^3}} \int{d {\bf k}
\over 2\omega (k)}\Bigg[ k^{2}(\sinh\!2g +\cosh\!2g)
+\omega^2 (k)(\cosh\!2g -\sinh\!2g)\Bigg]
\nonumber\\
&+& \frac{1}{2}m_\sigma^2I+6\lambda\sigma_0^2 I+3\lambda I^2
+\frac{1}{2}m_\sigma^2\sigma_0^2 +\lambda \sigma_0^4.
\label{en}
\end{eqnarray}
Extremising the above energy density with respect to the function $g(k)$ yields 
\begin{equation}
\tanh\!{2 g(k)}=-\,{{6 \lambda I+6 \lambda {\sigma _0}^2}\over {
{\omega (k)}^{2}+6 \lambda I+6 \lambda {\sigma _{0}}^{2}}}.
\label{gk}
\end{equation}
It is clear from the above equation that in the absence of a 
quartic coupling no such condensates are favoured since the
condensate function vanishes for $\lambda=0$.
Now substituting this value of $g(k)$ in the expression
for the $\sigma$-meson energy density yields 
\begin{equation}
\epsilon_\sigma=
{1 \over {2}}m_\sigma^2{\sigma _0}^2\,+\lambda {\sigma _0}^4
+\frac{1}{2}\frac{1}{(2\pi)^3}\int d{\bf k} (k^2+M_\sigma^2)^{1/2}
-3 \lambda I^2
\label{pot}
\end{equation}
where 
\begin{equation}
M_\sigma^2=m_\sigma^2+12\lambda I +12\lambda \sigma_0^2~,
\label{m2}
\end{equation}
with
\begin{equation}
I=\frac{1}{(2\pi)^3}\int\frac{d{\bf k}}{2} 
\frac{1}{({\bf k}^2+M_\sigma^2)^{1/2}}~,
\label{I}
\end{equation}
obtained from equation (\ref{gep9c}) after substituting for the condensate 
function $g(k)$ as in equation (\ref{gk}).
The expression for the ``effective potential" $\epsilon_\sigma$
contains divergent integrals. Since our approximation is 
nonperturbatively self-consistent, the field-dependent
effective mass $M_\sigma$ is also not well defined because
of the infinities in the integral $I$ given by equation (\ref{I}).
Therefore we first obtain a well-defined finite expression
for $M_\sigma$ by renormalisation. 
We use the renormalisation prescription of ref. \cite{politzer,pi}
and thus obtain 
the gap equation for the effective sigma meson mass, $M_\sigma$ 
in terms of the renormalised mass
and coupling parameters, $m_R$ and $\lambda_R$, as
\begin{equation}
M_\sigma^2=m_R^2+12\lambda_R\sigma_0^2+12\lambda_R I_f(M_\sigma),
\label{mm2}
\end{equation}
where
\begin{equation}
I_f(M_\sigma)=\frac{M_\sigma^2}{16\pi^2}
\ln \Big(\frac{M_\sigma^2}{m_R^2} \Big).
\label{if}
\end{equation}

After renormalisation \cite{mishra,ph4} and subtracting out
the pure vacuum contributions, the expressions for the
energy density and pressure can be written as

\begin{equation}
\epsilon=\sum_B \frac {\gamma_B}{2\pi^2} \int _0 ^{k_{FB}} 
(k^2+{M_B^*}^2)^{1/2}
k^2 dk +\Delta \epsilon_{ren} +\Delta \epsilon_\sigma
+\frac {1}{2} {m_\omega}^2 \omega_0^2
+\frac {1}{2} {m_\rho}^2 \rho_{03}^2,
\end{equation}
and
\begin{equation}
p= \frac {1}{3}\sum_B \frac {\gamma_B}{2\pi^2} \int _0 ^{k_{FB}} 
\frac {k^4}{(k^2+{M_B^*}^2)^{1/2}}
dk -\Delta \epsilon_{ren} -\Delta \epsilon_\sigma
+\frac {1}{2} {m_\omega}^2 \omega_0^2
+\frac {1}{2} {m_\rho}^2 \rho_{03}^2.
\end{equation}

In the above, $\Delta\epsilon_\sigma$ includes the quantum
effects from the scalar sigma meson and is given as \cite{mishra}

\begin{eqnarray}
\Delta \epsilon_\sigma 
&=& \frac{1}{2} m_R^2 \sigma_0^2+ 3\lambda_R \sigma_0^4 
+\frac {M_\sigma^4}{64\pi^2}
\Biggl(\ln\Big(\frac{M_\sigma^2}{m_R^2}\Big)-\frac{1}{2} \Biggr)
-3\lambda_R I_f^2\nonumber\\
&-&\frac {M^4_{\sigma,0}}{64\pi^2}
\Biggl(\ln\Big(\frac{M_{\sigma,0}^2}{m_R^2}\Big)-\frac{1}{2} \Biggr)
+3\lambda_R I_{f0}^2,
\label{vph0}
\end{eqnarray}
where $M_{\sigma,0}$ and $I_{f0}$ are the expressions
as given by eqs. (\ref{mm2}) and (\ref{if}) with $\sigma_0=0$.

The self consistency condition satisfied by $\sigma_0$, 
is now modified due to quantum effects from sigma field, 
from (\ref{sg0}), to

\begin{equation}
\frac {d(\Delta \epsilon_\sigma)}{d\sigma_0}
=-\sum_B g_{\sigma B} \frac {\gamma_B}{2\pi^2}
\int _0^{k_{FB}} \frac {k^2 dk M_B^*}{(k^2+{M_B^*}^2)^{1/2}}-
\frac {d \Delta\epsilon_{ren}}{d\sigma_0}
\label {sg}
\end{equation}

Explicitly, the left hand side of equation (\ref{sg}) is given as,
\begin{equation}
\frac {d(\Delta \epsilon_\sigma)}{d\sigma}=m_R^2\sigma_0
+12\lambda_R \sigma_0^3 +\frac {M_\sigma^2}{32 \pi^2}
\ln \Bigg (\frac{M_\sigma^2}{m_R^2}\Bigg)
\cdot \Bigg  ( \frac{d M_\sigma^2}{d\sigma_0}\Bigg)
-6\lambda_R I_f C_F  \Bigg  ( \frac{d M_\sigma^2}{d\sigma_0}\Bigg),
\end{equation}
where, $$\Bigg  ( \frac{d M_\sigma^2}{d\sigma_0}\Bigg)=
\frac {24\lambda_R \sigma_0}{1-12\lambda_R C_F},\;\;\;\;
C_F=\frac {1}{16\pi^2}\Bigg (1+\ln \Big (\frac{M_\sigma^2}{m_R^2}\Big)~.
\Bigg )$$

We might note here that apart from $\sigma_0$ satisfying the
selfconsistency condition (\ref{sg}), the in-medium scalar
meson mass, $M_\sigma$ also satisfies the selfconsistency
condition given by equations (\ref{mm2}) and (\ref{if}).
As earlier, the fermi momenta $k_{Fi}$'s are fixed from the baryon
and electric charge conservations given through (\ref{chrgb})
and (\ref{chrg0}) along with the selfconsistency conditions for the
meson fields given by (\ref{selfcons}). This then gives us the equation
of state for hyperonic matter, including the quantum correction
effects from the baryons and scalar mesons. We shall next consider the effects
of such an EOS on the neutron star structure.

\subsection{Neutron star structure}

The equations for the structure of a relativistic  spherical and 
static star composed of a perfect fluid were derived from Einstein's 
equations by Oppenheimer and Volkoff \cite{tov}. They are

\begin{mathletters}
\begin{equation}
\frac{dp}{dr}=-\frac{G}{r}\frac{\left[\epsilon+p\right ]\left[M+
4\pi r^3 p\right ]}{(r-2 GM)},
\label{tov1}
\end{equation}
\begin{equation}
\frac{dM}{dr}= 4\pi r^2 \epsilon ,
\label{tov2}
\end{equation}
\end{mathletters}
with $G$ as the gravitational constant and $M(r)$ as the enclosed gravitational 
mass. We have used $c=1$.
Given an EOS, these equations can be integrated from the origin as an initial
value problem for a given choice of central energy density, $(\epsilon_c)$.
The value of r (=R), where the pressure vanishes defines the
surface of the star.

We solve the above equations to study the structural properties of the
neutron star, using the EOS derived for the electrically charge
neutral hyperon rich dense matter including the quantum correction
effects from the baryons and scalar mesons. In an earlier study
\cite{prafulla}, the effect of vacuum strucure on neutron stars
was considered without inclusion of strange baryon sector.

\section {Results and Discussions}

We now proceed with the numerical calculations of
deriving the equation of state for hyperon rich dense matter
including the quantum correction effects from the
baryon and scalar meson sectors.
The nucleon meson couplings $C_s^2=g_{\sigma N}^2 M_N^2/m_\sigma^2$ 
and $C_v^2=g_{\omega N}^2 M_N^2/m_\omega^2$ are obtained from the
nuclear matter saturation properties of binding energy $-15.75$ MeV
and equilibrium nuclear matter density, $\rho_0$ of $0.193~{\mbox fm}^{-3}$.
The values of $C_s^2$ = 114.7, 167.5, 137.9 and $C_v^2$ = 183.3, 96.45,
63.7, for RHA and for $\lambda_R$=1.8 and 5 respectively
The compressibility of nuclear matter turns out to be 401 MeV
for $\lambda_R$ = 1.8, and 329 MeV for $\lambda_R$ = 5 \cite{mishra},
which are smaller than that of RHA. A similar
lowering of nuclear matter compressibilty was also observed 
in the modified RHA, where the renormalization scale
was chosen to be different from the free nucleon mass \cite{prakash}. 

The nucleon $\rho$ coupling,
$g_{\rho N}$ is determined from the symmetry energy coefficient,
$a_{sym}$, given as \cite {sym}
$$ a_{sym}=(\frac {g_{\rho N}}{m_\rho})^2 \cdot \frac {k_{F0}^3}{12\pi^2}
+\frac {k_{F0}^2}{6(k_{F0}^2+{M_N^*}^2)}$$
where $k_{F0}$ is the Fermi momentum of nuclear matter at saturation
density, $\rho_0$.
In our calculations, we choose $a_{sym}=32.5$ MeV, which gives
the value for the nucleon-rho meson coupling as $g_{\rho N}$ = 6.82,
7.0 and 7.25, for RHA and $\lambda_R$ = 1.8 and 5,  respectively.

The composition of the charge neutral dense matter depends
very sensitively on the hyperon-meson coupling  parameters,
which are, however, very poorly known \cite{prakash,glnd1}.
On general quark counting arguments we shall take the vector couplings 
from the $SU(6)$ theory, 
as $\frac {1}{3} g_{\omega N}=\frac {1}{2}
g_{\omega \Lambda}=\frac {1}{2} g_{\omega \Sigma}=g_{\omega \Xi}$
and, $g_{\rho N}=\frac {1}{2}g_{\rho \Sigma}=g_{\rho\Xi}$,
$g_{\rho\Lambda}=0$ \cite{ellis,debadesh,schaffner} and fix
the scalar couplings from the potential depth of the hyperon
in nuclear matter as, $U^N_\Lambda=U^N_\Sigma=-30$ MeV
and $U^N_\Xi=-28$ MeV \cite{schaffner}.

In Fig.1, we plot the equation of state, for RHA (equivalent to
considering the vacuum polarization effects from baryon sector)
and with quantum corrections
from both baryon and scalar meson sectors, for couplings
$\lambda_R~=~1.8$ and $\lambda_R~=~5$. Inclusion of hyperons softens
the EOS \cite{prakash} of the charge neutral dense matter. 
This is seen from the
change in slope of Fig. 1 for energy densities greater than $\approx$ 350
-500 MeV/fm$^3$ when hyperons start to appear. It is also observed that
quantum effects arising from scalar meson sector soften the equation of state
as compared to RHA and the softening increases for higher values of
the self coupling $\lambda_R$.

In Fig. 2, the effective masses for the baryons are plotted as
functions of the baryon number density.
All the in-medium baryon masses $M_B^*=M_B+g_{\sigma B}\sigma_0$
decrease with density. The variation for hyperon masses is slower as
compared to nucleon masses. This a reflection of a smaller coupling
of hyperons to the sigma field. Further, we note that the decrease in the
effective mass, $M_B^*$ with density, is slower when
quantum corrections from the sigma mesons through condensates are taken into 
account. The value of $M_B^*$ increases with $\lambda_R$ which is a 
reflection of diminishing baryon sigma coupling strength for larger
value of quartic self-interaction.

The meson field strengths are plotted in Fig. 3. The potentials
can be obtained by multiplying the corresponding meson couplings
to the baryons. The vector potential initially rises linearly
and then the rise is slowed down for densities $\sim 2 \rho_0$,
when the hyperons start to appear \cite {schaffner}. The magnitude
as well as the rate of rise is decreased with inclusion of
quantum corrections from the $\sigma$ mesons. Such a
decrease is also observed for the scalar potential, which
leads to a smaller value for the anti-baryon potential,
$(g_{\omega B} \omega_0-g_{\sigma B}\sigma_0)$, due to the
$\sigma$ meson quantum correction effects. The scalar field
tends to saturate at high densities. The magnitude of
isovector vector field, $\rho_{03}$ remains small (reaching
a maximum of $\sim$ 15-20 MeV) throughout the density regime.

Inclusion of quantum corrections from the sigma meson leads to a 
self-consistent calculation of the in-medium sigma mass, given by
equation (\ref{mm2}) which we plot in figure 4. There is an 
increase in the scalar meson mass with density, since
the quartic coupling $\lambda_R$ is positive. The increase
is sharper for higher value of the quartic coupling \cite {mishra}.

In figures 5, 6 and 7, the fractional particle densities, $\rho_i/\rho$ 
are plotted for the situations with and without the quantum corrections
from scalar mesons.
At $\rho/\rho_0 \sim 2$, first $\Lambda$, the lightest hyperon 
appears in all these situations.
It is noticed that for little higher value for the density,
the negatively charged hyperon
$\Xi^-$ appears, which competes with the leptons in maintaining
charge neutrality. Because of the larger coupling to the 
vector meson as compared to that of $\Sigma$-hyperon,
the effective chemical potential for $\Xi^-$ becomes large
enough to have the threshold for occurrence of $\Xi^-$
earlier than $\Sigma^-$. We might note here that, this is 
in line with the extrapolated $\Sigma^-$ atomic data \cite
{atomic}, which suggest that $\Xi^-$ may feel repulsion at
high density, which would mitigate against its appearance 
in dense matter \cite{glnd1}, although this remains inconclusive.
The appearance of $\Xi^-$ leads to a depletion
of the lepton concentrations consistent with charge
neutrality condition. Such a depletion is also seen, 
for example in the magnitude of the electron chemical potential, 
$\mu_e=-\mu_E$, plotted in figure 8. The $\Xi^-$-hyperon becomes 
more populous than $\Lambda$ almost immediately after it starts appearing.
The proton concentration increases and saturates at higher densities, 
and the appearance of $\Sigma^+$ maintains charge neutrality condition.
We might note here that the threshold condition for a given
species to appear, is given by the equation (\ref{thresh}).
Thus the order in which they appear is  rather sensitive to the 
hyperon-meson coupling constants. With the couplings chosen 
in the present paper, the threshold density for $\Sigma^+$ 
is smaller than $\Sigma^0$ and $\Sigma^-$. This is due to the fact
that first of all $\Sigma^+$ has a lower mass and its
coupling to the iso-vector vector, $\rho$- meson decided by
isospin leads to a higher effective chemical potential,
compared to $\Sigma^0$ or $\Sigma^-$. We also notice that
after an initial rise of $\Sigma^+$ partial density,
it starts decreasing at a higher density when $\Xi^-$ 
saturates, to maintain charge neutrality as electron
density keeps decreasing in this density regime.
Inclusion of quantum effects from scalar meson
shifts the threshold densities for the hyperons to higher
values. This is due to the fact that such quantum corrections
lead to a higher effective mass as illustrated in Fig. 2,
and hence the threshold condition (\ref{thresh}) is met 
at a higher density. 

In figure 9, we plot the mass of the neutron star as a function
of the central energy densities for the situations
of with and without the quantum correction effects
from scalar mesons. For subnuclear densities, we use the 
results of Negele and Vautherin \cite{negele} for the energy 
density regime 0.025 MeV/fm$^3~ <~\epsilon~<$ 50 MeV/fm$^3$, 
the EOS of Baym, Pethick and Sutherland \cite {baym} for 
5$\times$ 10$^{-10}$ MeV/fm$^3~<~\epsilon~<$ 0.025 MeV/fm$^3$, and
the EOS of Feynman {\it et al} \cite{feynman} below 10$^{-10}$ MeV/fm$^3$.
The value of the maximum mass of the neutron star in solar mass units is
found to be 1.717, 1.6 and 1.42 for RHA, $\lambda_R$ = 1.8 and 5
respectively, along with the corresponding radii as
12.0, 11.62 and 11.1 kilometers.  
Thus it is seen that inclusion of quantum effects reduces
the maximum mass of stable neutron stars,
which reflects the fact that the EOS becomes softer due to
quantum effects, as illustrated in figure 1. Such a decrease
of maximum mass for the neutron star due to quantum effects
was also seen in Ref. \cite{prafulla}. 
In fig. 10, the radius of the neutron star versus its
mass is plotted for the different situations.
The quantum effects are seen to reduce the radius of the maximum
mass of the neutron star.
\section{summary}
To summarize, in the present work, we have derived the EOS
of charge neutral hyperonic matter including the quantum
effects in a nonperturbative manner. The method consists of
having an explicit ansatz for the ground state of dense
matter with baryon-antibaryon condensates as well as
scalar meson condensates. Extremisation of the thermodynamic
potential leads to determination of the condensate functions
in a self consistent manner. This gave rise to a softer
EOS. The vacuum polarization effects through baryon-antibaryon condensates
give rise to the Relativistic Hartree approximation
of summing over the tadpole diagrams for the baryon propagator.
Inclusion of quartic self interaction with
a positive coupling for the $\sigma$ meson enabled us
to include the quantum effects arising from the $\sigma$ field.
However, we have not considered here the quantum effects 
arising from the $\omega$ and $\rho$ mesons.
The composition of the charge neutral dense matter is
known to be rather sensitive to the hyperon meson
coupling parameters, which involve large uncertainties.
In particular, the uncertainty of the coupling
of $\Sigma^-$ in dense matter raises the question of the behaviour 
of the electrochemical potential in the neutron star matter,
which is crucial in the possible presence of the kaon condensed
phase in the core of the neutron stars\cite{glnd1}. With inclusion of quantum
effects from $\sigma$ meson, one notices that the threshold densities
for the onset of appearance of the hyperons increase, due to the
larger values of the respective effective masses.
We have also considered the neutron star structural properties
using such an equation of state, which leads to a smaller value
for the maximum mass as compared to when quantum effects
from scalar mesons are not taken into account.
The approximation here lies in the specific ansatz for the
ground state structure. However, it would be interesting
to investigate with a more general ground state including 
$\omega$ and $\rho$ mesons than what has been considered
here. Work in this and related fields are in progress.

\section{Acknowledgement}
One of the authors (P. K. P.) would like to acknowledge 
FAPESP (Processo-99/08544-0) for financial support and
the IFT, S\~ao Paulo, for kind hospitality.

\begin{figure}
\begin{center}
\psfig{file=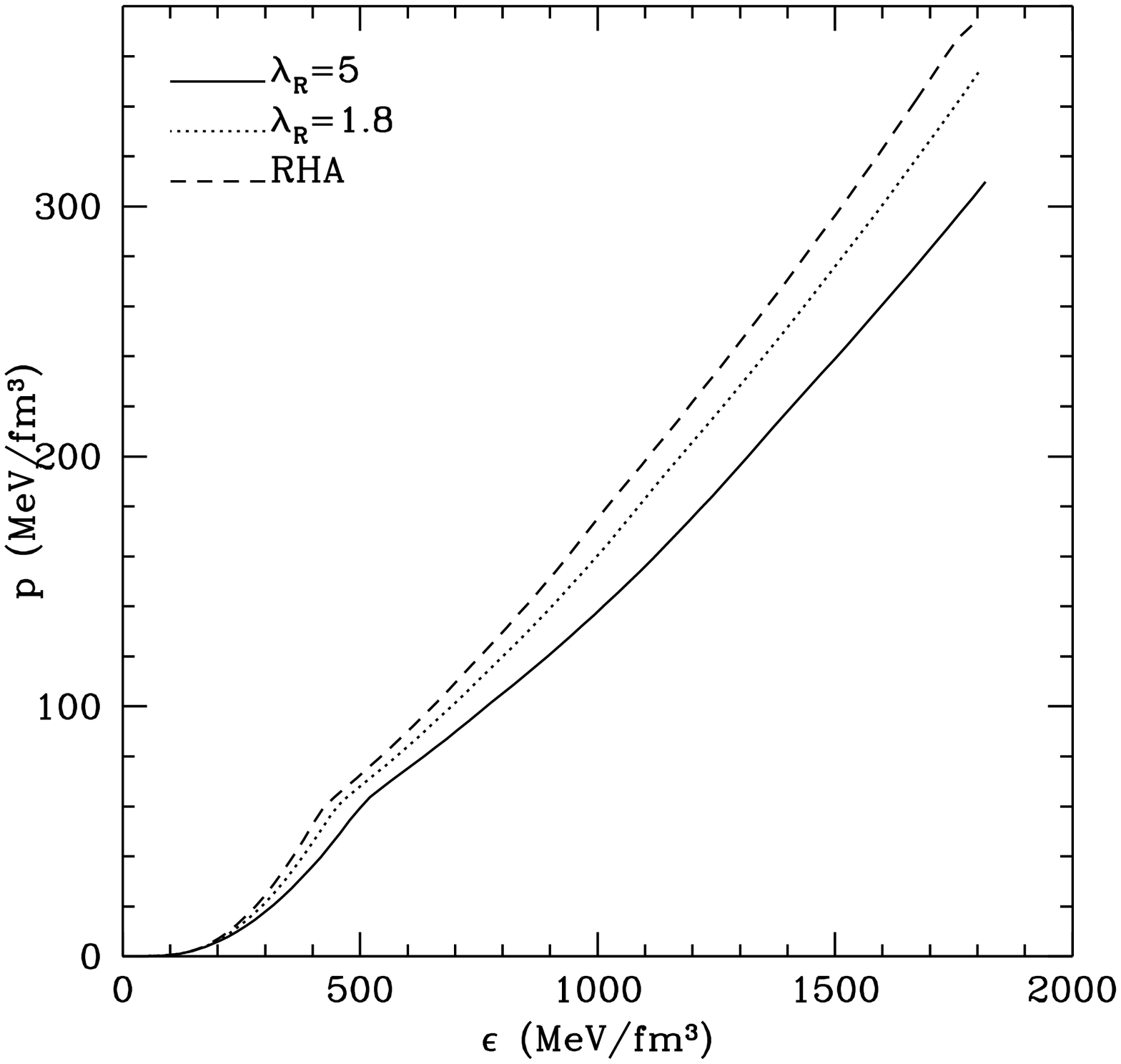,width=9cm,height=9cm}
\end{center}
\caption{The equation of state for hyperonic matter for RHA, $\lambda_R=1.8$
and 5.0. The quantum correction from scalar mesons gives a softer equation 
of state.}
\end{figure}
\begin{figure}
\begin{center}
\psfig{file=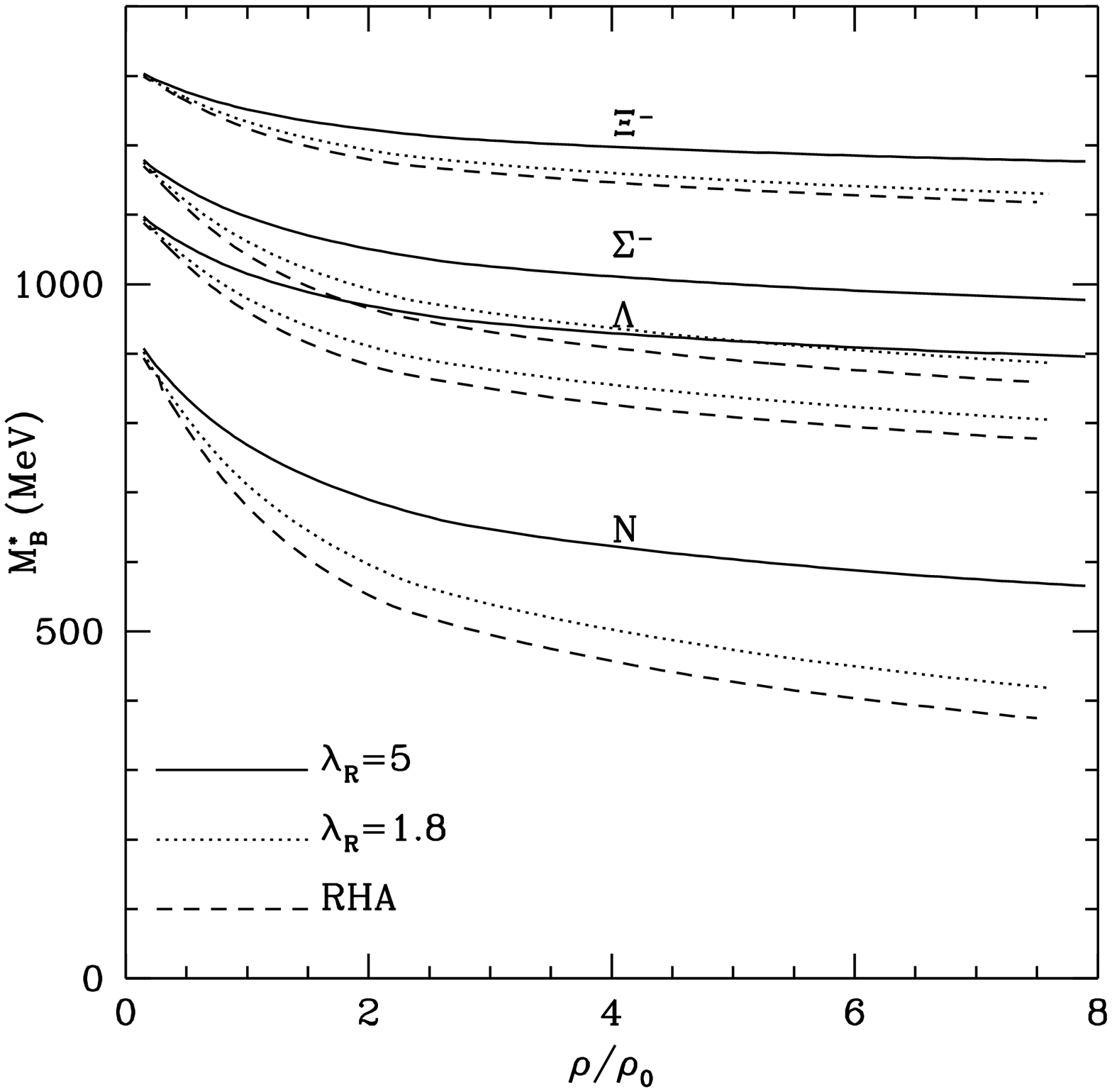,width=9cm,height=9cm}
\end{center}
\caption{Effective baryon masses in the medium.}
\end{figure}
\begin{figure}
\psfig{file=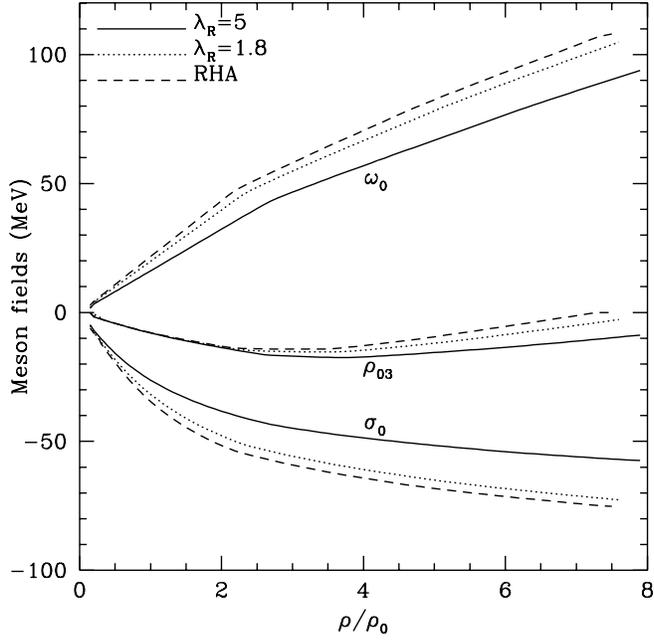,width=9cm,height=9cm}
\caption {The meson field strengths as functions of baryon density.}
\end{figure}
\begin{figure}
\psfig{file=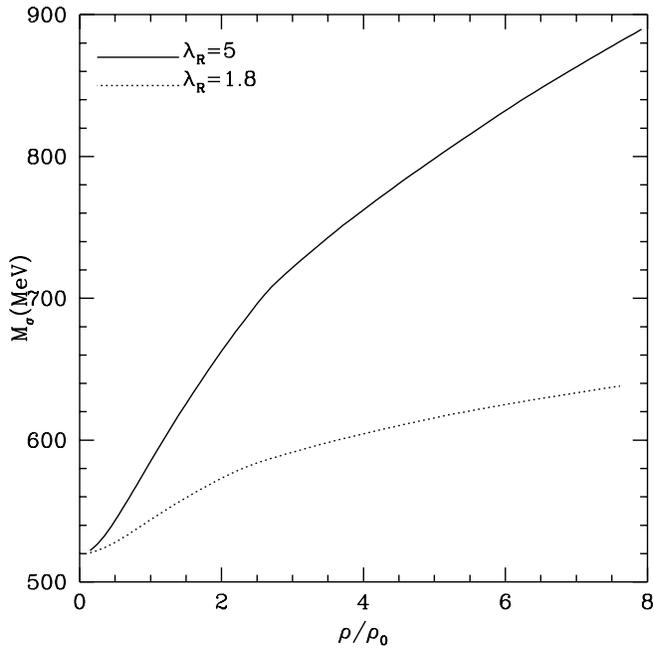,width=9cm,height=9cm}
\caption{In medium scalar meson mass versus baryon density.}
\end{figure}
\begin{figure}
\psfig{file=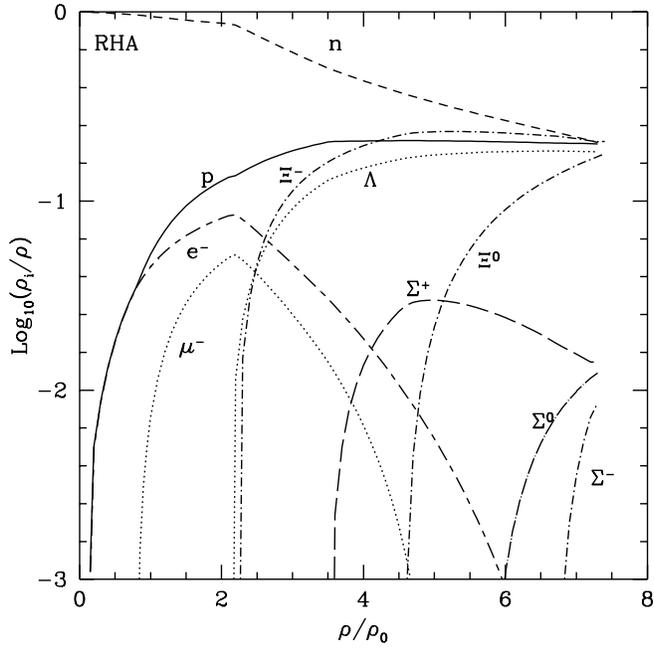,width=9cm,height=9cm}
\caption{Particle density fractions in Relativistic Hartree Approximation.}
\end{figure}
\begin{figure}
\psfig{file=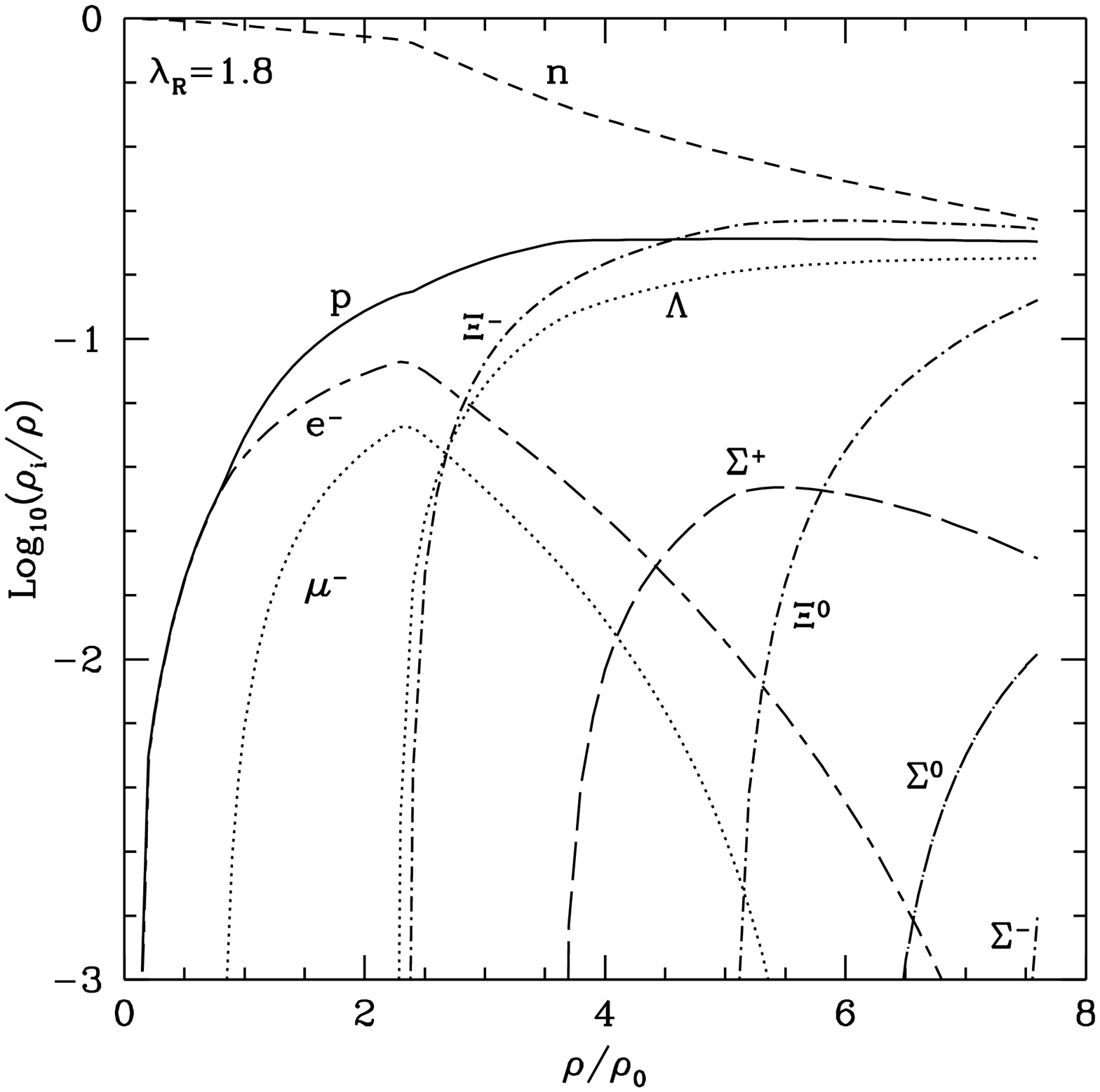,width=9cm,height=9cm}
\caption{Same as fig. 5 with quantum corrections from scalar mesons for
$\lambda_R=1.8$.}
\end{figure}
\begin{figure}
\psfig{file=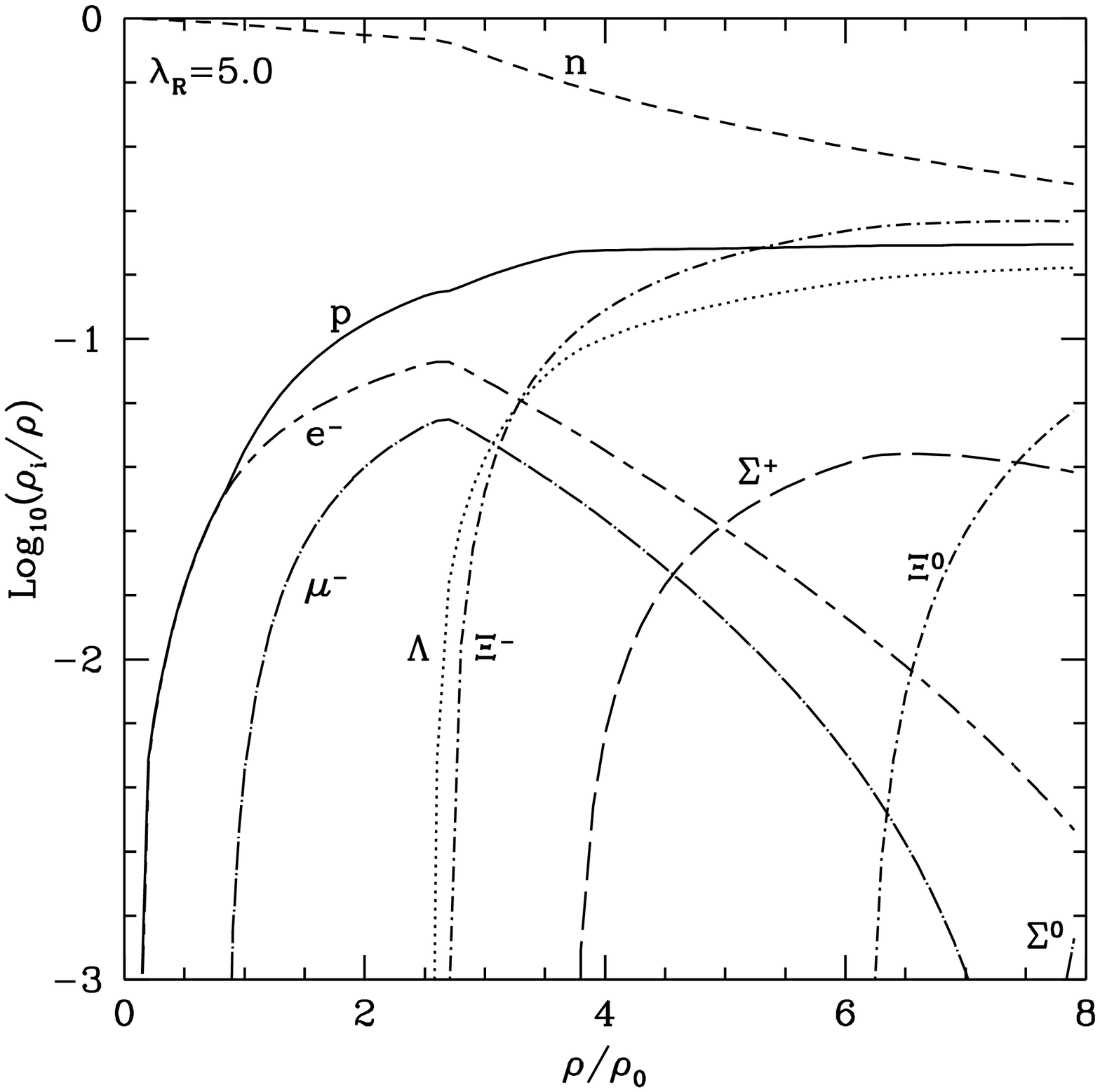,width=9cm,height=9cm}
\caption{Same as fig. 5 with quantum corrections from scalar mesons for
$\lambda_R=5$.}
\end{figure}
\begin{figure}
\psfig{file=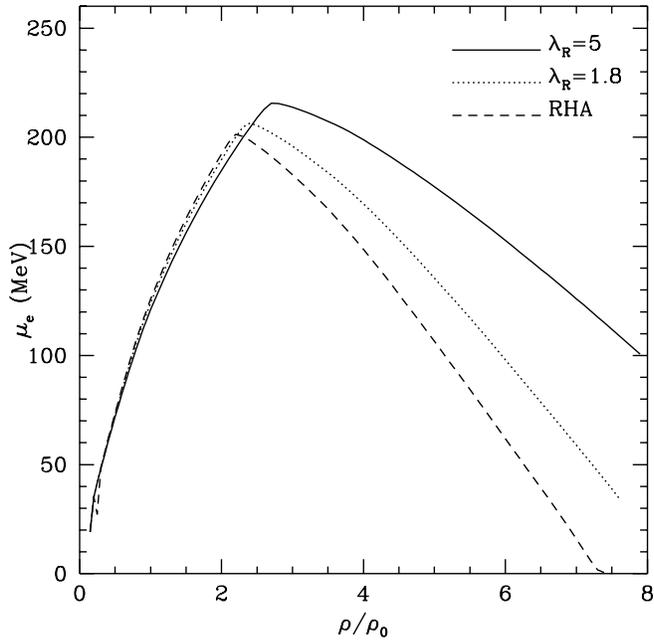,width=9cm,height=9cm}
\caption{Electron chemical potential versus density.}
\end{figure}
\begin{figure}
\psfig{file=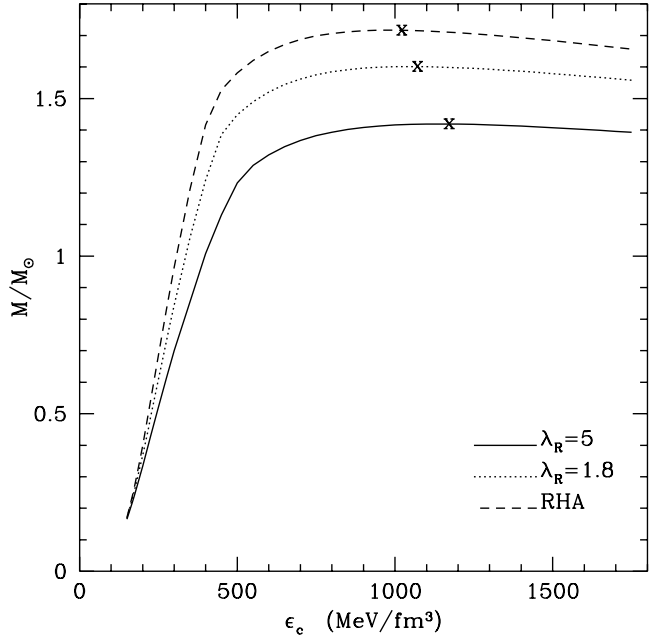,width=9cm,height=9cm}
\caption{Mass of neutron star versus central energy density, $\epsilon_c$. 
The crosses indicate the maximum masses.}
\end{figure}
\begin{figure}
\psfig{file=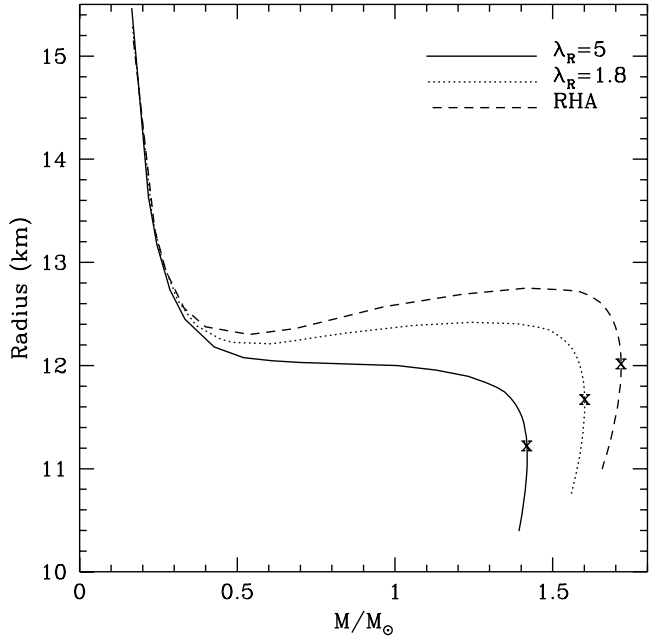,width=9cm,height=9cm}
\caption{The radius versus mass of the neutron star for RHA, $\lambda_R=1.8$,
and 5.0. The radius of the star decreases with the quantum correction}
\end{figure}
\end{document}